\documentclass[a4paper]{jpconf}
\usepackage{graphicx}

\usepackage{amsmath}
\usepackage{amssymb}
\usepackage{color}

\begin{document}

\title{Probing new physics with atmospheric neutrinos at KM3NeT-ORCA}

\author{Jo\~{a}o A. B. Coelho for the KM3NeT collaboration}

\address{APC, Universit\'{e} Paris Diderot, CNRS/IN2P3, Sorbonne Paris Cit\'{e}, F-75205 Paris, France}

\ead{jcoelho@apc.in2p3.fr}

\begin{abstract}
We present the prospects of ORCA searches for new physics phenomena using atmospheric neutrinos. Focus is given to exploiting the impact of strong matter effects on the oscillation of atmospheric neutrinos in light of expanded models, such as sterile neutrinos and non-standard interactions. In the presence of light sterile neutrinos that mix with active neutrinos, additional resonances and suppressions may occur at different energies. One may also use neutrino oscillations to probe the properties of the coherent forward scattering which may be altered by new interactions beyond the Standard Model. Preliminary studies show that ORCA would be able to probe some parameters of these models with sensitivity up to one order of magnitude better than current constraints. 
\end{abstract}

\section{Atmospheric Neutrinos in KM3NeT-ORCA}
\label{intro}

ORCA is one of the detectors in the KM3NeT project which are currently under construction in the Mediterranean Sea with the goals of measuring the Neutrino Mass Hierarchy (NMH), and searching for astrophysical high energy neutrino sources \cite{LoI}. The ORCA detector will be densely instrumented with 2070 optical modules separated from each other by 9 m vertically and 20m horizontally, to enable the study of atmospheric neutrinos ranging from a few GeV to $\sim$100 GeV. In this energy range, refraction in the matter of the earth significantly alters the effective neutrino flavour mixing. These matter effects can generate resonant neutrino flavour transitions which are the main driver of ORCA's sensitivity to the NMH. The impact of strong matter effects may also be exploited to search for new physics phenomena. With an effective mass of 5.7~Mton, ORCA will provide unprecedented statistical power in the energy range relevant for the study of resonant neutrino transitions. 

\begin{figure}[ht]
\begin{center}
\begin{minipage}{16pc}
\includegraphics[width=16pc]{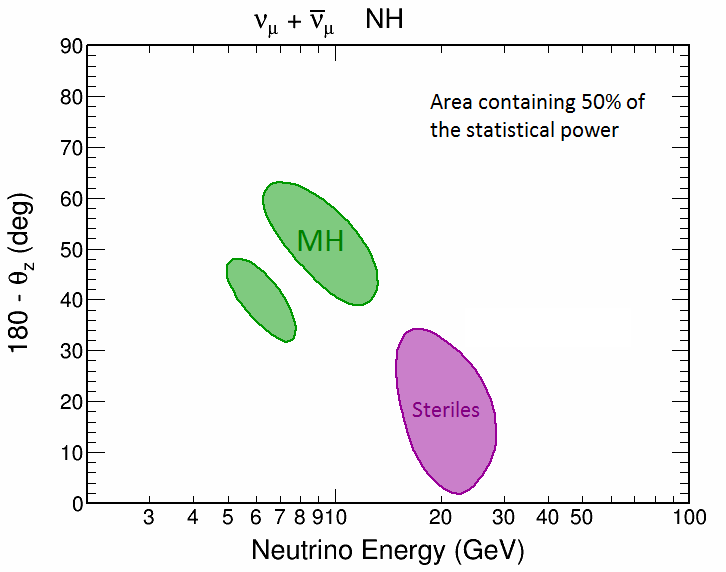}
\end{minipage} \hspace{2pc}%
\begin{minipage}{16pc}
\includegraphics[width=16pc]{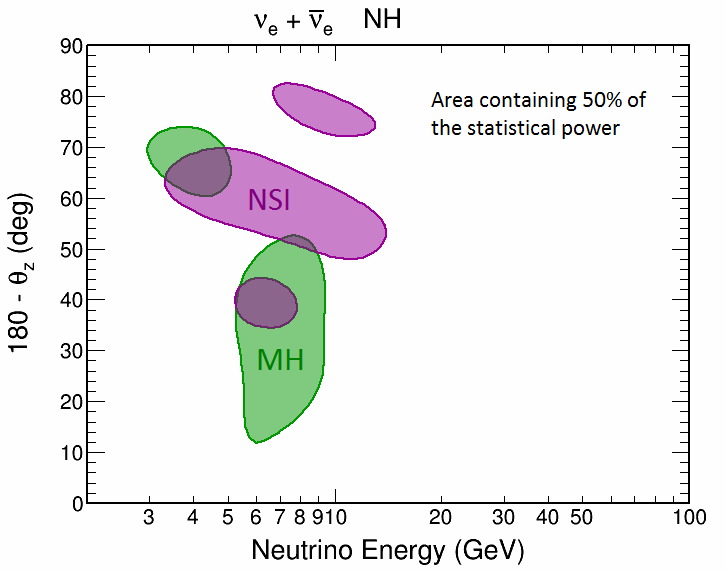}
\end{minipage}\\
\begin{minipage}{32pc}
\caption{\label{fig:phase_space}Regions of parameter space where 50\% of the sensitivity to new physics is expected to be found (Purple) compared to where the sensitivity to the NMH is expected to be found (Green).}
\end{minipage}
\end{center}
\end{figure}

The main method used in ORCA for measuring the NMH is the measurement of the atmospheric neutrino rate as a function of energy and direction. Distortions in the oscillation pattern due to matter effects are expected to occur in specific parts of this energy and direction phase space. Atmospheric neutrinos provide a wide range of energies and baselines (equivalent to direction), which can be used to constrain systematic effects in the regions where matter effects are not expected to occur. This same methodology can be employed to also search for physics beyond the Standard Model, by looking for distortions characteristic of resonant effects in particular models. Since these resonant effects often occur in different parts of the energy-direction phase space, these models can be studied independently of the standard matter effects. Figure \ref{fig:phase_space} shows the regions of this phase space where most of the signal of the NMH, sterile neutrinos, and NSI are expected to appear.

\section{Extended models}

The new physics models we will study here are phenomenological extensions of the standard neutrino oscillation picture. In the case of NSI, the interaction Hamiltonian of the neutrino system is modified into an arbitrary unitary form motivated by the potential existence of new neutrino interactions that behave as four-fermion point interactions at low energies. The extended NSI Hamiltonian is shown in equation \ref{eq:nsi}, with $U$ being the PMNS matrix, $G_F$ is Fermi's constant and $n_e$ is the electron density in the medium.

\begin{equation}
\label{eq:nsi}
H_{NSI}=U\left(\begin{array}{ccc}0&0&0\\0&\frac{\Delta{m^2_{21}}}{2E}&0\\0&0&\frac{\Delta{m^2_{31}}}{2E}\end{array}\right)U^\dag\pm\sqrt{2}G_Fn_e\left(\begin{array}{ccc}1+\epsilon_{ee}&\epsilon_{e\mu}&\epsilon_{e\tau}\\\epsilon_{e\mu}^*&\epsilon_{\mu\mu}&\epsilon_{\mu\tau}\\\epsilon_{e\tau}^*&\epsilon_{\mu\tau}^*&\epsilon_{\tau\tau}\end{array}\right)
\end{equation}

In the sterile neutrino case, the Hamiltonian is extended to four neutrinos. The 4th neutrino flavour is assumed to not interact in any way, yielding an effective potential difference between the sterile and active flavours due to Neutral Current (NC) interactions. Equation \ref{eq:sterile} shows the resulting Hamiltonian form.

\begin{equation}
\label{eq:sterile}
H_{sterile}=U\left(\begin{array}{cccc}0&0&0&0\\0&\frac{\Delta{m^2_{21}}}{2E}&0&0\\0&0&\frac{\Delta{m^2_{31}}}{2E}&0\\0&0&0&\frac{\Delta{m^2_{41}}}{2E}\end{array}\right)U^\dag\pm\sqrt{2}G_F\left(\begin{array}{cccc}n_e&0&0&0\\0&0&0&0\\0&0&0&0\\0&0&0&n_n\end{array}\right)
\end{equation}
 
The NC interaction is effectively proportional to the density of neutrons in the medium ($n_n$). Detailed reviews of these models can be found in references \cite{nsi} and \cite{sterile}.

\section{Preliminary sensitivities}

In this poster, we presented initial preliminary sensitivities to some NSI and sterile neutrino parameters under simplified assumptions. First, detector related effects such as energy and angular resolution, as well as detection efficiency, were incorporated in a parametrised fashion, which does not take into account all possible correlations between these effects. Second, the identification of neutrino flavours was assumed to be perfect. Finally, no systematic effects were considered and all nuisance oscillation parameters were kept fixed at their global best fit values. These assumptions are in general optimistic and a full realistic analysis is currently under development.

Nevertheless, the sensitivities presented here are promising, given that they represent only one year of ORCA data and preliminary studies have demonstrated selection efficiencies of $\sim$70\% for muon neutrinos and $\sim$90\% for electron neutrinos. These efficiencies are also expected to improve in future.

Figure \ref{fig:phase_space} shows the results of the preliminary sensitivities to sterile neutrinos in the $|U_{\mu4}|^2-|U_{\tau4}|^2$ parameter space and to NSI in the $\epsilon_{\tau\tau}-\epsilon_{e\tau}$ parameter space. The sensitivities obtained improve on current limits by an order of magnitude in the NSI parameters and about a factor of 5 in the $|U_{\tau4}|$ mixing. In the case of sterile neutrinos, sensitivities were computed for $\Delta{m^2_{41}}=0.3~\mbox{eV}^2$, but the sensitivity is expected to be valid for any $\Delta{m^2_{41}}\gtrsim0.1~\mbox{eV}^2$ 

\begin{figure}[ht]
\begin{center}
\begin{minipage}{16pc}
\includegraphics[width=16pc]{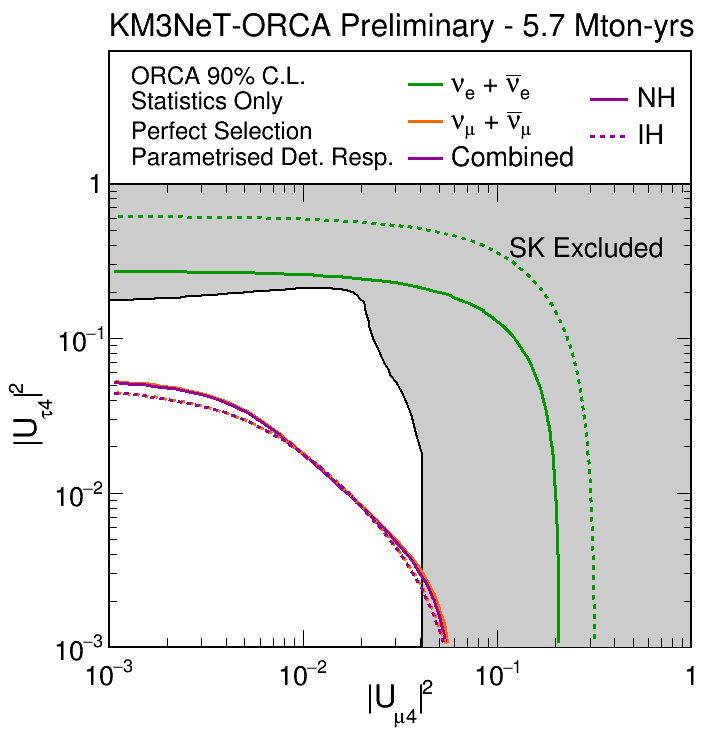}
\end{minipage} \hspace{2pc}%
\begin{minipage}{16pc}
\includegraphics[width=16pc]{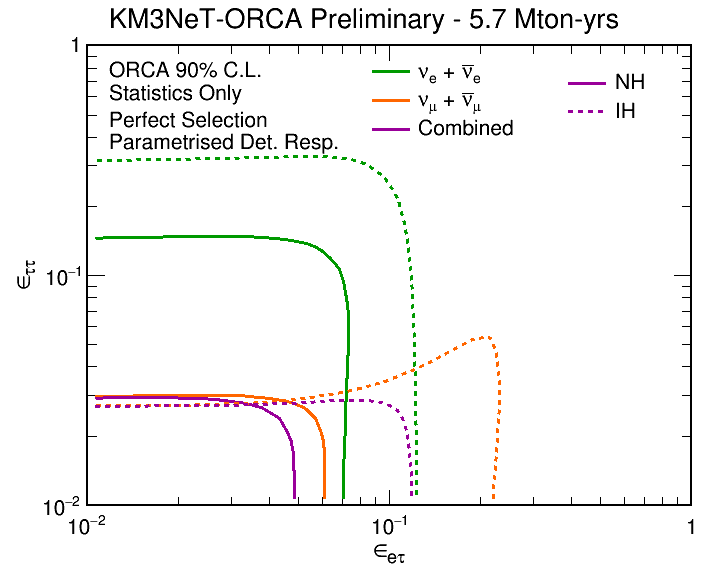}
\end{minipage}\\
\begin{minipage}{32pc}
\caption{\label{fig:contours}Preliminary sensitivity of the ORCA detector to sterile neutrinos (Left) and NSI (Right) in selected slices of the multi-dimensional parameter space of the models. The sensitivities were computed for the equivalent of 1 year of ORCA data and simplified optimistic assumptions were made. For comparison, limits from the Super-Kamiokande experiment \cite{superk} are shown for sterile neutrinos.}
\end{minipage}
\end{center}
\end{figure}

\section{Summary}

In summary, we studied the potential of ORCA to measure effects of new physics with atmospheric neutrinos. We focused on two extensions of the neutrino oscillation framework, sterile neutrinos and NSI, in which matter effects are expected to play a significant role. Preliminary results are encouraging, showing potential improvements of an order of magnitude in the sensitivity to some parameters in these models in comparison with current limits.

\section*{Acknowledgments}

This work was supported by IdEx and LabEx UnivEarthS programs at Sorbonne Paris Cit\'e (ANR-11-IDEX-0005-02, ANR-10-LABX-0023)\\[6pt]

\end{document}